\documentstyle[12pt,epsf,epsfig]{article}
\newcommand{\uq}{{\sf u}}                  
\newcommand{\sq}{{\sf s}}                  
\newcommand{\qq}{{\sf q}}                  
\newcommand{\qqbar}{{\sf \=q}}             
\newcommand{\gq}{{\sf g}}                  

\begin{document}

\baselineskip=18truept

\date{\normalsize \today}

\title{\Large \bf Gluon Polarization from correlated
high-$p_T$ hadron pairs in polarized electro-production}

\author{Alessandro Bravar$^\ast$, Dietrich von Harrach, and
Aram Kotzinian$^\dag$ \\
\\
\normalsize Institut f\"ur Kernphysik, Universit\"at Mainz,
D-55099 Mainz, Germany \\}

\maketitle

\vspace*{20mm}

\begin{abstract}
\baselineskip=18truept
We propose to measure the polarized gluon distribution function
$\Delta G (\eta)$ at $\eta \stackrel{<}{\sim} 0.2$ via {\it pseudo} jet
production in polarized fixed target lepton nucleon scattering
at typical lepton beam energies of 200~GeV.
The measurement of the spin-asymmetry for the production of correlated
charge conjugated hadrons of opposite transverse momentum
can be directly related to $\Delta G (\eta)$
through the photon-gluon fusion process.
We also present a numerical analysis of the accuracy which can be 
obtained for different flavors and kinematics
of the observed hadron pair. 
\end{abstract}

\vspace*{\fill}

$^\ast$ corresponding author: a.bravar@cern.ch

$^\dag$ On leave from Yerevan Physics Institute,
375036 Yerevan, Armenia, and JINR, 141980 Dubna, Russia 

\newpage

\section{Introduction}

Polarized deep inelastic scattering (DIS) experiments have shown
that the quark spins account for only a rather small fraction of
the nucleon spin~\cite{emc}, thus implying an appreciable contribution
either from gluon spins or possibly from orbital angular momentum.
Competing explanations exist for this result, in which the
polarized glue $\Delta G$ or the negatively polarized
strange quarks $\Delta {\sf s}$ lower the quark contribution
to the nucleon spin. One way to solve this puzzle is to measure 
$\Delta G$ directly
by studying, for example, polarized semi-inclusive processes,
where the gluons enter in the initial state of the hard scattering
sub-processes at lowest order in $\alpha_s$.

\begin{figure}
\vspace{-10mm}
\begin{center}
\mbox{
\epsfxsize=16cm\epsffile{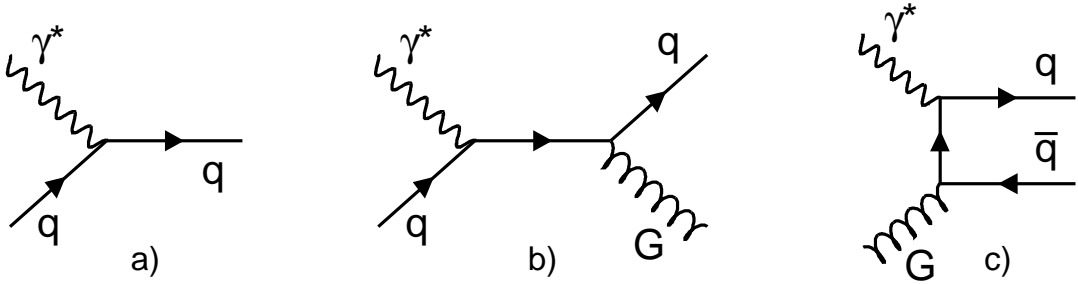}
}
\end{center}
\vspace{-5mm}
\caption{ 
Lowest order Feynman diagrams for DIS $\gamma^* N$ scattering:
a) virtual photo-absorbtion, b) gluon radiation (Compton diagram),
c) photon-gluon fusion (PGF).
}
\label{fig:feydia}
\end{figure}

Since the main hard scattering sub-process in DIS is the virtual
photo-absorption $\gamma^\ast q \rightarrow q$
(\qq-event, fig.~\ref{fig:feydia}a),
the measurement of inclusive polarized structure functions,
like $g_1(x)$, does not allow the separation of the various parton components,
and does not give direct access to the spin-dependent gluon distribution
$\Delta G (\eta)$.
In principle, the integral $\Gamma = \int \Delta G (\eta) \, {\rm d} \eta$
can be extracted indirectly from the QCD
analysis of the $Q^2$ dependence of $g_1(x,Q^2)$~\cite{bfr}.
Information on the shape of $\Delta G$, however, can only be obtained
from a direct measurement, which is in any case highly desirable.
At first order in $\alpha_s$ two hard sub-processes, 
the gluon radiation (\qq\gq-event, fig.~\ref{fig:feydia}b)
and the photon-gluon fusion (\qq\qqbar-event, fig.~\ref{fig:feydia}c)
contribute to the DIS cross section.
These QCD effects are not only clearly visible as $(2+1)$ jets at the
Hera Collider~\cite{hjets}, but also in the event shapes and
transverse momentum spectra of hadrons produced in DIS at typical 
fixed target energies of $100 - 500~{\rm GeV}$~\cite{e665}. 
The unpolarized gluon distribution has been also extracted in a
fixed target DIS experiment from the event shape analysis
of the events~\cite{e665g}.
 
Favorable conditions to measure $\Delta G (\eta)$ are given, for example,
in heavy flavor production~\cite{Glu88}
which proceeds via photon-gluon fusion (PGF),
and in the reaction
$\gamma^\ast \, + \, N \rightarrow 2$ high-$p_T$ jets $+ \, X$~\cite{Fel96}
which also proceeds via PGF (fig.~\ref{fig:feydia}c),
however with a non-negligible background from gluon radiation
(fig.~\ref{fig:feydia}b).
The latter measurement requires the detection of two jets with large
transverse momenta ($p_T \, ({\rm jet}) > 5~{\rm GeV}/c$).
At the moderate energies of fixed target experiments the 
criteria for identifying jets are not unambiguous
(this is the case also at collider energies)
due to their large angular spread and low particle multiplicity.
On the other hand, leading hadrons produced in DIS reflect the original
parton direction and flavor.
The detection of these hadrons or {\it pseudo} jets allows
to tag the partons emerging from the hard sub-processes of
figs.~\ref{fig:feydia}b and~\ref{fig:feydia}c.
This is also valid for fixed target experiments
with typical incident lepton energies of 200~GeV.

We propose to look for two correlated high-$p_T$ hadrons,
$h_1$ and $h_2$, in the forward hemisphere ($x_F > 0$)
with $p_T (h_1) > p_{T,min}$ and $h_2$
opposite in azimuth to $h_1$ with $p_T (h_2) > p_{T,min}$.
The measured cross section spin-asymmetry for $h_1 + h_2$ production
can be related to the gluon polarization $\Delta G (\eta)$.
For high enough $p_T$'s soft contributions to the $h_1 + h_2$ cross section
are small, so only the hard part
(figs.~\ref{fig:feydia}b and~\ref{fig:feydia}c)
contributes significantly to the cross section.
From the study of this process a $p_{T,min} > 1.0 - 1.5 ~{\rm GeV}/c$
is found to be sufficiently large to suppress almost completely the 
contribution of the leading order sub-process (fig.~\ref{fig:feydia}a)
to the $h_1 + h_2$ cross section.
We will also consider the production of such hadron pairs 
in the photo-production limit
using the full virtual photon spectrum down to quasi-real photons.

When developing these ideas we have found that similar arguments
have been proposed previously~\cite{Str95} but not to the extent given here
and without a quantitative analysis and without predictions.
Proposals to measure $\Delta G$ by detecting a single high-$p_T$ hadron
accompanied by a second fast hadron with a large longitudinal momentum
have been also made~\cite{Fon81}.
In this case a very large $p_T$ cut-off of about 3 to 4~GeV/$c$ is required
to select the processes of figs.~\ref{fig:feydia}b and~\ref{fig:feydia}c
which considerably reduces the event yields.

\section{Electro-production of high-$p_T$ hadrons}

Factorization allows the decomposition of the (polarized) electro-production
cross section up to the first nontrivial order in pQCD
for the production of high-$p_T$ hadron pairs ($h_1 + h_2$) as
\setcounter{equation}{0}
\renewcommand{\theequation}{1\alph{equation}}
\begin{eqnarray}
(\Delta)\sigma^{h_1 + h_2} = & \sum_q \left\{ (\Delta)q \otimes
(\Delta){\hat \sigma}^{\gamma^\ast q \rightarrow q} 
\otimes D_q^{h_1 + h_2} \right\} + \\
  & \sum_q \left\{ (\Delta)q \otimes
(\Delta){\hat \sigma}^{\gamma^\ast q \rightarrow qG} 
\otimes D_{q,G}^{h_1 + h_2} \right\} + \\
  & \sum_q \left\{ (\Delta)G \otimes
(\Delta){\hat \sigma}^{\gamma^\ast G \rightarrow q\bar{q}} 
\otimes D_{q,\bar{q}}^{h_1 + h_2} \right\} \; .
\end{eqnarray}
\setcounter{equation}{1}
\renewcommand{\theequation}{\arabic{equation}}

\noindent The decomposition reflects the processes
shown in fig.~\ref{fig:feydia}.
The convolutions ($\otimes$) are performed over the hard
scattering kinematics, the (polarized) quark $(\Delta)q$
and gluon $(\Delta)G$ distributions,
and the fragmentation functions involved.
The sums $\sum_q$ run over the (anti)quark flavors.
$(\Delta){\hat \sigma}^{\gamma^\ast q \rightarrow q}$, 
$(\Delta){\hat \sigma}^{\gamma^\ast q \rightarrow qG}$, and
$(\Delta){\hat \sigma}^{\gamma^\ast G \rightarrow q\bar{q}}$ are
the (polarized) partonic hard-scattering cross sections
for the zero order \qq-event,
and the first order $\alpha_s$ \qq\gq- and \qq\qqbar-events,
respectively~\footnote{At present,
the unpolarized cross sections are calculated to next to leading order,
while the polarized ones to leading order only.
For their exact leading order expressions see ~\protect\cite{poldis,Fie89}.}.
$D_q$, $D_{q,G}$, and $D_{q,\bar{q}}$ describe the fragmentation of partons
to $h_1 + h_2$.
As long as the polarization of the final hadronic state is not studied
they are spin-independent.
For more details see for instance~\cite{Fie89}.

The cross sections in eqn.~1 exhibit collinear divergencies 
since the $\alpha_s$ order QCD amplitudes for
${\hat \sigma}^{\gamma^\ast q \rightarrow qg}$ and
${\hat \sigma}^{\gamma^\ast g \rightarrow q\bar{q}}$
are divergent in the $t$ channel the first
($\cos \hat{\vartheta} \rightarrow -1$, where $\hat{\vartheta}$ is the c.m.
angle between the incoming $\gamma^\ast$ and the outgoing quark)
and in the $t$ and $u$ channels the latter
($\cos \hat{\vartheta} \rightarrow \mp 1$).
To avoid these singularities in a Monte Carlo simulation procedure
some cut-offs on the matrix elements are usually imposed.
The choice, however, is not unique~\cite{KMS89}.
In our case the selection of correlated {\it pseudo} jets with large $p_T$'s
implies that most of the c.m. energy of the hard sub-process $\sqrt{\hat{s}}$ goes to the outgoing parton transverse momenta $\hat{p}_T$,
and that the outgoing partons are produced close to
$\hat{\vartheta} = \pm 90^\circ$ in the c.m.,
therefore far away from the kinematical regions
with these collinear divergencies,
where also the higher order corrections are presumably larger.
This criterion is basically equivalent to the $z-\hat{s}$ scheme
used to regulate the divergencies in the matrix elements~\cite{KMS89}.
The most natural scales for the studied process are the transverse
momenta $\hat{p}_T$ of the outgoing partons in the c.m..
Typically, they are chosen to be $\sum \hat{p}_T^2$ and varied between
$\frac{1}{2}$ and 2 times this value.

The parton distribution functions (PDF's) in 
\qq\gq- and \qq\qqbar-events are probed at a momentum fraction
\begin{equation}
\eta = (\hat{s} + Q^2)/2 M E_{\gamma^\ast} = x_{Bj} \, (\hat{s}/Q^2 +1)
\label{eq:eta}
\end{equation}
where $E_{\gamma^\ast}$ is the energy of the virtual photon,
and $M$ the nucleon mass (note that $\eta \geq x_{Bj}$).
With a typical lepton beam energy of 200~GeV 
these PDF's can be accessed in the region $\eta > 0.02$.
As far as the spin of the nucleon is concerned, that appears to be 
the most interesting region, since the largest contribution to 
$\Gamma = \int \Delta G(\eta) {\rm d} \eta$
is expected to come from this region~\cite{bfr}.
The proposed method will probe the PDF's at a scale
$\stackrel{>}{\sim} 10~{\rm GeV}^2$.

\begin{figure}
\vspace*{-5mm}
\begin{center}
\mbox{
\epsfysize=6cm\epsffile{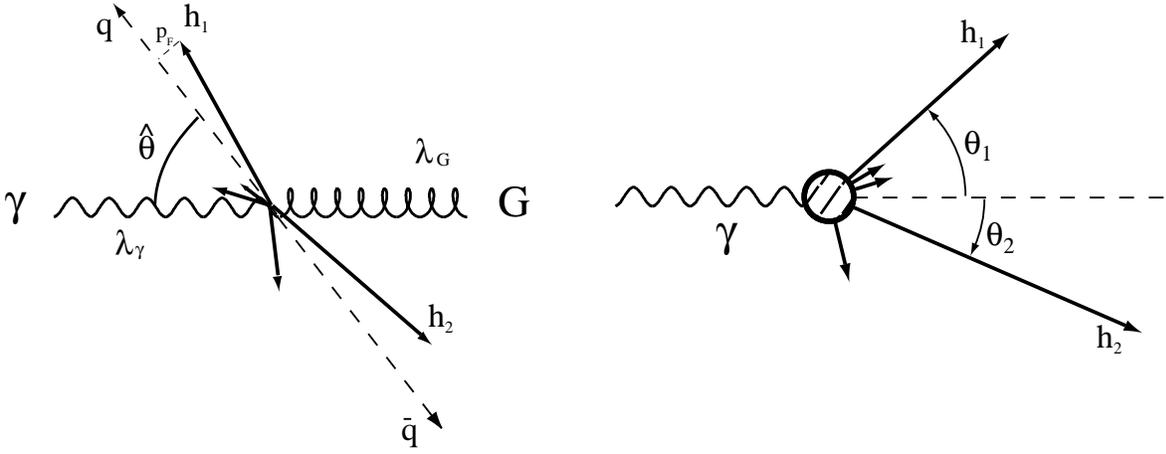}
}
\end{center}
\vspace*{-5mm}
\caption{ 
Correlated high-$p_T$ hadron production as seen in the 
$\gamma^\ast$ - parton
c.m. (left) and in the laboratory (right).
The two frames are related by a Lorentz bost with
$\gamma = E_{\gamma^\ast} / \protect\sqrt{\hat{s}}$.
}
\label{fig:frag}
\end{figure}

Next we examine the fragmentation process to high $p_T$ hadrons
(relative to the virtual photon axis).
This is a non-perturbative process for which we have no detailed 
predictions and at present can be tackled 
only via a model approach by using, for instance,
the observed hadron spectra to parametrize
the fragmentation functions (FF's).
In the fragmentation process hadrons acquire some
transverse momentum, $p_F$, with respect to the initial quark direction,
which in the LUND string model is usually parametrized with a gaussian
distribution with $\langle p_F \rangle \sim 0.36~{\rm GeV}/c$
and a harder tail at the one percent level~\cite{And83}.
Hadrons acquire also part of their transverse momentum $p_T$ from 
the intrinsic transverse momentum, $k_T$, of the partons in the nucleon
($p_T \sim z \cdot k_T$, where $z$ is the fraction of the parton energy
carried by the hadron).
The intrinsic $k_T$ is also parametrized with a gaussian distribution with
$\langle k_T \rangle \sim 0.44~{\rm GeV}/c$.
For hadrons emerging from \qq-events these are the only sources of
transverse momentum, and the resulting $p_T$ drops off exponentially with
$\langle p_T^2 \rangle \sim
\langle p_F^2 \rangle + z^2 \langle k_T^2 \rangle$.
The experimental data at low $p_T$ are well reproduced
by this description~\cite{pt}.
For example, the fraction of \qq-events with at least one hadron with
$p_T > 1.5~{\rm GeV}/c$ is of the order $< 10^{-3}$.

In contrast to this scenario, the $p_T$ spectra of hadrons emerging
from events with an underlying first order QCD sub-process show a
power law dependence and hence fall off much more slowly
as seen in the experimental data~\cite{pt}.
In this case a considerable fraction of $p_T$ comes from the
transverse momentum $\hat{p}_T$ of the outgoing partons, and
the intrinsic $p_F$ and $k_T$ will have only a minor effect
on the $p_T$'s of the leading hadrons,
provided that the $p_T$ is sufficiently large.
So the direction of these hadrons in the partonic c.m. will approximately be
that of the fragmenting partons (see fig.~\ref{fig:frag}).
A cut-off of $p_{T,min} > 2~{\rm GeV}/c$ should be already sufficient
to start suppressing the \qq-events in the single hadron $p_T$ spectra.
Higher values for $p_{T,min}$ in excess of $3 - 4~{\rm GeV}/c$
would be preferable.
Such a high value for the $p_T$ cut-off, however, reduces significantly
the event yields.
For two high-$p_T$ hadrons opposite in azimuth, instead,
a considerably lower $p_{T,min}$ cut-off is sufficient
to suppress the \qq-events.
If $\hat{p}_T$ is large enough ($\hat{p}_T > 2~{\rm GeV}/c$),
one would expect that the two outgoing partons
will fragment independently into hadrons.
In this case the FF's $D_{q,G}^{h_1 + h_2}$
and $D_{q,\bar{q}}^{h_1 + h_2}$ to two high-$p_T$ hadrons
in eqns.~1b and~1c represent indeed a product of two one-particle FF's:
$D_{q,G}^{h_1 + h_2} = D_q^{h_1} \times D_G^{h_2} + (1 \leftrightarrow 2)$ and
$D_{q,\bar{q}}^{h_1 + h_2} = 
D_q^{h_1} \times D_{\bar{q}}^{h_2} + (1 \leftrightarrow 2)$.

\section{The Asymmetry and $\Delta G / G$}

The polarized electro-production cross section spin-asymmetry
$A_{LL}^{lN \rightarrow h_1 h_2}$ at order $\alpha_s$ can be written as
\begin{equation}
A_{LL}^{lN \rightarrow h_1 h_2} =  
\frac{\sum_q \left\{ \Delta G \otimes \Delta 
               \sigma^{\gamma^\ast G \rightarrow q \bar{q}}
               \otimes D_{q,\bar{q}}^{h_1 + h_2} \right\} +
      \sum_q \left\{ \Delta q \otimes \Delta 
               \sigma^{\gamma^\ast q \rightarrow q G}
               \otimes D_{q,G}^{h_1 + h_2} \right\}}
     {\sum_q \left\{ G \otimes        
               \sigma^{\gamma^\ast G \rightarrow q \bar{q}}
               \otimes D_{q,\bar{q}}^{h_1 + h_2} \right\} +
      \sum_q \left\{ q \otimes        
               \sigma^{\gamma^\ast q \rightarrow q G}
               \otimes D_{q,G}^{h_1 + h_2} \right\}} \; .
\label{eq:all}
\end{equation}
The notation is as in eqn.~1.
The virtual photon depolarization with respect to the incident muon
polarization has been absorbed into the polarized cross sections.
The contribution to $A_{LL}^{lN \rightarrow h_1 h_2}$
from the leading order diagram
(fig.~\ref{fig:feydia}a) is strongly suppressed when selecting high-$p_T$
correlated hadron pairs (see next section),
and therefore it has not been included in eqn.~\ref{eq:all}.
On the other hand, the spin-asymmetry for the leading order sub-process
is very small in the relevant kinematical range,
and at most would introduce a small dilution in
$A_{LL}^{lN \rightarrow h_1 h_2}$.

\begin{figure}
\vspace{-10mm}
\begin{center}
\mbox{
\epsfxsize=16cm\epsffile{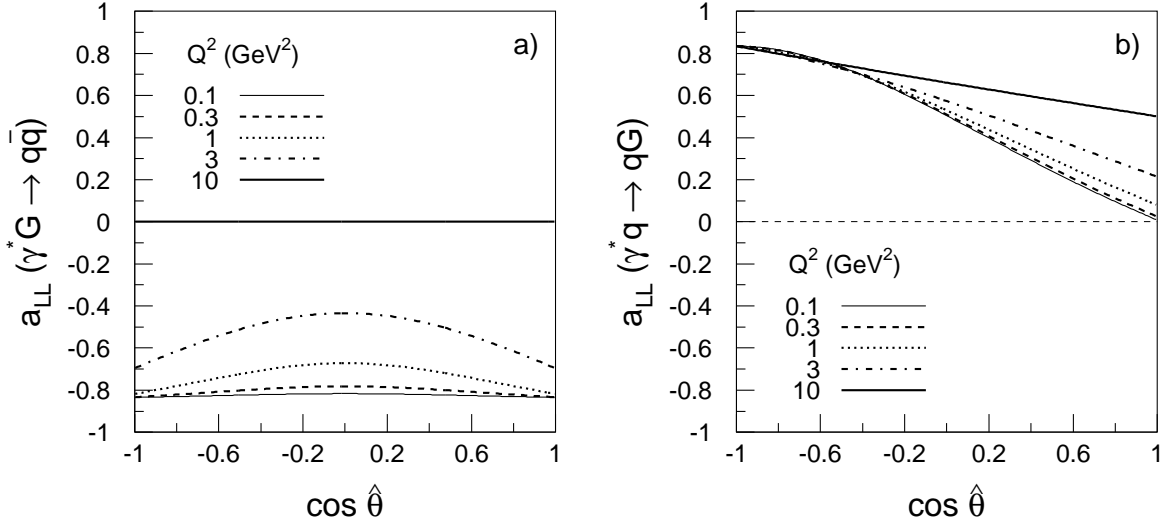}
}
\end{center}
\vspace{-10mm}
\caption{ 
Basic scattering asymmetries $\hat{a}_{LL}$ for the photon-gluon fusion (a)
and the gluon radiation (b)
as a function of the c.m. scattering angle $\hat{\vartheta}$
for different values of $Q^2$
at fixed $\hat{s} = 10~{\rm GeV}^2$ and fixed $y = 0.7$ ($D \approx 0.83$).
Note their $Q^2$ dependence.
}
\label{fig:asymm}
\end{figure}

Two competing sub-processes of the same order in $\alpha_s$ contribute
to the (polarized) cross section and $A_{LL}^{lN \rightarrow h_1 h_2}$.
Only the PGF, however, is of interest for the extraction
of the (polarized) gluon distribution $(\Delta) G (\eta)$,
while the Compton process acts as a background.
To quantify this contribution we introduce the ratio
\begin{equation}
R = \frac{\sigma^{PGF}}{\sigma^{COMPT}} =
    \frac{\sum_q \left\{ G \otimes
            \hat{\sigma}^{\gamma^\ast G \rightarrow q \bar{q}} 
            \otimes D_{q,\bar{q}}^{h_1 + h_2} \right\} }
    {\sum_q q \left\{ \otimes \hat{\sigma}^{\gamma^\ast q \rightarrow q G}
            \otimes D_{q,G}^{h_1 + h_2} \right\} } \propto
    \frac{G(\eta)}{F_1(\eta)} \; ,
\label{eq:rrr}
\end{equation}
which depends in particular on the covered $\eta$ region.
The Compton background is proporitonal to the unpolarized structure function
$F_1$, and over the covered kinematical region
it is dominated by scattering off \uq-{\it valence} quarks
because of their large charge compared to other quark flavors
and large density compared to {\it sea} quarks.

For a qualitative discussion of the asymmetry 
$A_{LL}^{lN \rightarrow h_1 h_2}$, 
showing explicitly the contributions of different processes,
eqn.~\ref{eq:all} can be approximated as
\begin{equation}
A_{LL}^{lN \rightarrow h_1 h_2} \approx 
\langle \hat{a}_{LL}^{\gamma^\ast G \rightarrow q \bar{q}}\rangle
\frac{\Delta G}{G} \, \frac{R}{1+R} + 
\langle\hat{a}_{LL}^{\gamma^\ast q \rightarrow q G}\rangle \, A_1 \, 
\frac{1}{1+R} \;.
\label{eq:all2}
\end{equation}
$A_1$ is the virtual photo-absorbtion asymmetry, and
the $\langle\hat{a}_{LL}\rangle$'s are the scattering asymmetries
at partonic level ($\hat{a}_{LL} = \Delta\hat{\sigma} / \hat{\sigma}$).
At present they are calculated to order $\alpha_s$ only~\cite{poldis}.
Figure~\ref{fig:asymm} shows these asymmetries for the two sub-processes 
considered above as a function of the c.m. scattering angle $\hat{\vartheta}$
for different values of the photon virtuality $Q^2$
at fixed $\hat{s} = 10~{\rm GeV}^2$ and fixed $y = 0.7$
($y \approx E_{\gamma^\ast} / E_{beam}$) to which
corresponds a depolarization $D$ of about 83~\%.
For most of the proposed measurement the PGF asymmetry
$\hat{a}_{LL}^{\gamma^\ast G \rightarrow q \bar{q}}$ (fig.~\ref{fig:asymm}a)
is close to $-1 \times D$.
The asymmetry $\hat{a}_{LL}^{\gamma^\ast q \rightarrow q G}$ for the
Compton process (fig.~\ref{fig:asymm}b) instead is always positive
and close to $+0.5 \times D$ for scattering at $90^\circ$ in the c.m..
The two sub-processes, therefore, contribute with opposite signs and
weights to $A_{LL}^{lN \rightarrow h_1 h_2}$.
Since $|\langle\hat{a}_{LL}^{\gamma^\ast G \rightarrow q \bar{q}}\rangle|$
is about 2 times as large as
$|\langle\hat{a}_{LL}^{\gamma^\ast q \rightarrow q G}\rangle|$,
and the asymmetries of the relevant backgrounds are proportional to $A_1$,
which is rather small in the relevant range~\cite{emc},
this process offers favorable conditions for the measurement of
$\Delta G (\eta)$.

In order to extract the gluon polarization from the measured asymmetry
$A_{LL}^{lN \rightarrow h_1 h_2}$ one assumes that the leading order 
contribution is negligible and subtracts the contribution
of the Compton sub-process (fig.~\ref{fig:feydia}b) from eqn.~\ref{eq:all}.
The quark polarizations $\Delta {\sf q}$  over the relevant range
are already available from other measurements~\cite{Ade96}.
These backgrounds (proportional to $A_1$),
reduce to a dilution effect when using isoscalar targets,
because for such target materials $A_1$ is very small (close to zero)
in the relevant kinematical range~\cite{emc}.
The ratio $R = \sigma^{PGF} / \sigma^{COMPT}$ can be estimated 
from simulations of this process, but not measured directly.
The incomplete knowledge of $R$ will introduce the largest uncertainty 
in the extraction of $\Delta G / G$.
Therefore it is important to maximize this ratio by selecting,
for instance, specific configurations in the final state 
as discussed in the next section and
get reliable and stable predictions for it.

\section{Tagging the PGF}

We have studied the electro-production of two high-$p_T$ hadrons 
described above using existing event generators~\cite{lepto,ariadne,pythia},
which contain all the relevant underlying physics processes including
the exact treatment of the $\alpha_s$ first order matrix elements
(fig.~\ref{fig:feydia})
and are known to reproduce fairly accurately the final hadronic state
in a variety of processes, including the observed $p_T$ and $p_L$
hadron spectra, the angular distributions, the measured cross sections, etc..
We performed our quantitative analysis in the kinematical conditions of the
COMPASS experiment~\cite{comp} for a 200~GeV/$c$ $\mu^+$ beam
incident on a deuteron (iso-scalar) target with
$0.5 < y < 0.9$ ($100 < E_{\gamma^\ast} < 180~{\rm GeV}$) and
$W^2 > 200~{\rm GeV}^2$.
A slightly modified version of the electro-production LEPTO~\cite{lepto}
event generator has been used down to 
$Q^2 = 0.4~{\rm GeV}^2$ with the leading order unpolarized parton densities
of~\cite{GRV95}.
Different descriptions of the fragmentation process within the LUND
model~\cite{And83,pythia}
(string fragmentation or independent fragmentation) have been considered 
including the color dipole model of~\cite{ariadne}.
They all give results consistent with each other.
We also varied several settings in the generators in order to verify
the stability of the results.

The following selections of the high-$p_T$ hadron pairs, 
based on the discussions in the previous section
and the results of the simulations, are found to enhance the
relative contributions of the \qq\qqbar- and \qq\gq-events 
over the \qq-events.

\noindent {\bf A} -- $p_T$ {\it cut:}
There should be two hadrons in the event with $p_T > 1.0 - 1.5~{\rm GeV}/c$.
Since $\sqrt{\hat{s}} > m(h_1, h_2)$, where $m(h_1, h_2)$
is the invariant mass of the $h_1 + h_2$ pair,
we also demand that $m(h_1, h_2) > 2.5 - 3.0~{\rm GeV}/c^2$
in order to impose a lower limit on the c.m. energy $\sqrt{\hat{s}}$.
The final event yields depend sensitively on these cuts;
they will be chosen to obtain the
best compromise of statistical and systematical accuracies.

\noindent {\bf B} -- $x_F$ {\it cut:}
To avoid fragmentation effects from the target remnant
only hadrons produced in the forward hemisphere with
$x_F>0$ should be considered, which corresponds to a cut 
$z > 0.1$ for each selected hadron.

\noindent {\bf C} -- $\Delta \phi$ {\it cut:}
The two hadrons should be found opposite in azimuth,
such that $|\Delta \phi | = 180^\circ \pm 30^\circ$,
where $\phi$ is the azimuthal angle between
the lepton scattering plane and outgoing hadron.
Conservation of $p_T$ already introduces a strong correlation, if primordial
$k_T$ and higher order $\alpha_s$ contributions are small.
 
\begin{figure}
\vspace{-5mm}
\begin{center}
\mbox{
\epsfxsize=16cm\epsffile{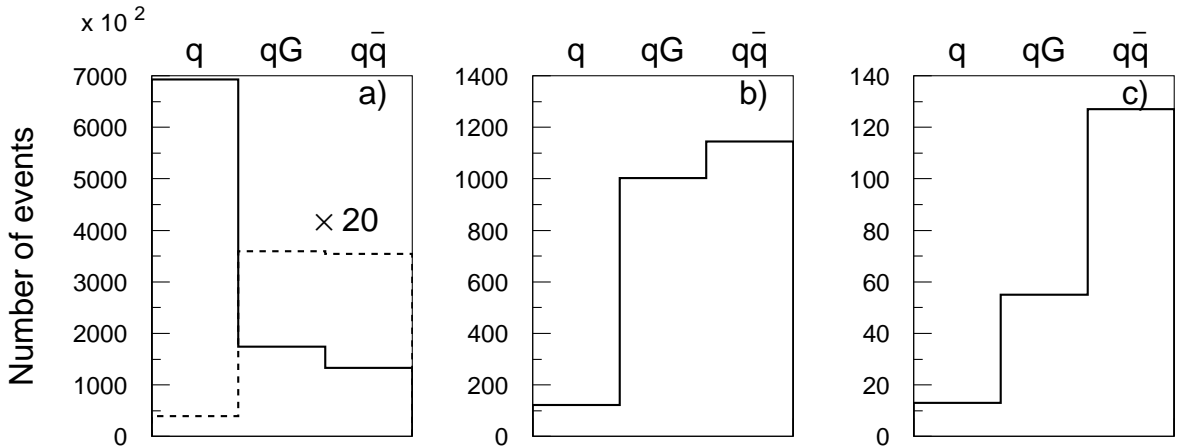}
}
\end{center}
\vspace{-5mm}
\caption{ 
Contributions of different sub-processes to the cross section:
(a) no cuts (full line) and selected hadron pairs
(cuts {\bf A} to {\bf C}, dashed line);
(b) oppositely charged hadron pairs;
(c) oppositely charged kaon pairs.
The event yields are normalized to $10^6$ generated events.
}
\label{fig:yields}
\end{figure}

As can be observed from fig.~\ref{fig:yields}a,
these selections are already strong enough to enhance the relative
contributions of the \qq\qqbar- and \qq\gq-events over the \qq-events
(dashed vs. full line).
Further, the relative contribution of the PGF over the Compton sub-process
can be enhanced as follows.

\noindent {\bf D} -- {\it oppositely charged hadrons:}
Fragmenting partons in \qq\qqbar-events have opposite charges
and favored fragmentation will preferentially lead to oppositely charged
leading hadrons in contrast
to \qq\gq-events, where the gluon fragments with equal probability
to positive and negative hadrons.
Therefore selecting events with oppositely
charged hadrons will enhance the contribution of the \qq\qqbar-events
over the \qq\gq-events.

\noindent {\bf E} -- {\it $K^+K^-$ pairs:}
The production of strange hadrons in fragmentation
is suppressed compared to non-strange hadrons, unless there is
already an {\sf s}~quark in the initial state.
This effect is parametrized with a strangeness suppression factor $\gamma_s$,
which presently ranges between 0.2 and 0.3~\cite{str}.
The production of a high $p_T$ correlated kaon pair will be 
therefore strongly suppressed,
unless there is already a fragmenting ${\sf s}{\sf \bar{s}}$ pair.
This selection, however, reduces the event yields by about a factor of 10.

\begin{figure}
\vspace{-5mm}
\begin{center}
\mbox{
\epsfxsize=16cm\epsffile{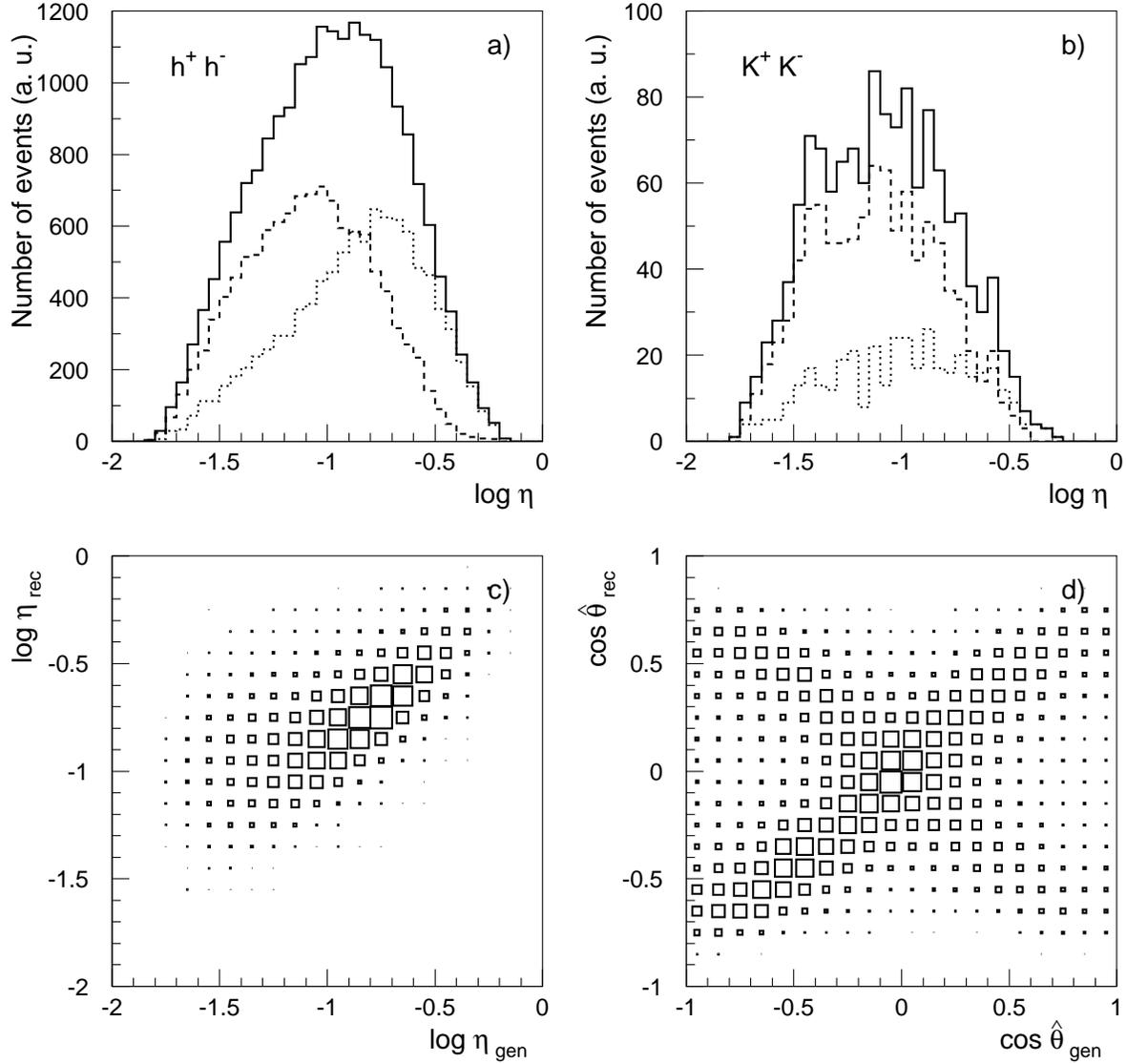}
}
\end{center}
\vspace{-5mm}
\caption{ 
$\eta$ distributions of the gluons (\qq\qqbar-events dashed line)
and of the quarks (\qq\gq-events dotted line) for the selected
high-$p_T$ hadron pairs (a) and kaon pairs (b).
The full line is the sum of the two. 
Correlations between the generated and the reconstructed 
parton momentum fraction $\eta$ (c)
and the scattering c.m. angle $\cos \hat{\vartheta}$ (d)
using the kinematics of the two selected hadrons only.
}
\label{fig:gluon}
\end{figure}

Figures~\ref{fig:yields}b and~\ref{fig:yields}c show the relative suppression
of the \qq-events for oppositely charged hadron and kaon pairs~\footnote{A 
hadron pair can consist also of hadrons of different flavors. The 
sample of kaon pairs forms a subset in the sample of hadron pairs.}, 
respectively, compared to fig.~\ref{fig:yields}a (full line)
where no selections on the final hadronic state have been applied.
The selections for the hadron pairs shown in fig.~\ref{fig:yields}b are
{\bf A} to {\bf D} and for the kaon pairs of fig.~\ref{fig:yields}c 
{\bf A} to {\bf E} with $p_T > 1.0~{\rm GeV}/c$ and
$m(h^+,h^-) > 2.5~{\rm GeV}/c^2$.
We will use the same cuts also in the following.
The fraction of \qq-events in both selected samples is around 5~\%.
The relative contribution of the \qq\qqbar-events over the \qq\gq-events
$R$ (eqn.~\ref{eq:rrr}) over the whole covered kinematical range
is about 1 for the selected $h^+ + h^-$ pairs
and about 2 for the selected $K^+ + K^-$ pairs
(see also figs.~\ref{fig:gluon}a and~\ref{fig:gluon}b
from where the $\eta$ dependence of $R$ can be inferred).
The reduction factors $r$ of the event rates are
$r^{h^+h^-} \sim 2.4 \times 10^{-3}$ and
$r^{K^+K^-}\sim 2.2 \times 10^{-4}$, respectively.
The large event yields in the $h^+ + h^-$ channel also allow one to apply
tighter selections for this sample such as $p_T > 1.5~{\rm GeV}/c$,
for which $r^{h^+h^-} \sim 2.6 \times 10^{-4}$.
These results have been obtained with the LEPTO program~\cite{lepto}
with the kinematical settings as discussed above.

We have also considered the production of the correlated high-$p_T$
hadron pairs in the photo-production limit,
because the total cross section, and correspondingly the event yields,
are substantially increased in the $Q^2 \rightarrow 0$ limit.
As already mentioned, the relevant scales for the studied process
are set by the scale of the hard sub-processes $\sum \hat{p}_T^2$,
and are around or above 10~GeV$^2$.
We have studied the photo-production limit with the photo-production
event generator PYTHIA~\cite{pythia}.
We have obtained results very close to the ones given above
for the electro-production case with $Q^2 > 0.4~{\rm GeV}^2$
($Q^2 > 1~{\rm GeV}^2$).
Also the contribution from non-pointlike photons has been found to be small.
We therefore assume the same results to hold over the $Q^2$ region
not covered with the simulation extending from $Q^2 = 0.4~{\rm GeV}^2$ 
($Q^2 > 1~{\rm GeV}^2$) to
$Q^2_{min} = m_\mu^2 \frac{E_{\gamma^\ast}^2}
{E_{beam}(E_{beam}-E_{\gamma^\ast})}$,
where $Q^2_{min}$
is determined from the energy-momentum conservation at the lepton vertex.
The estimated cross section for producing such hadron pairs 
$\sigma^{l N \rightarrow h^+h^-}$ ($\sigma^{l N \rightarrow K^+K^-}$)
is around $(150 \pm 50)~{\rm nb} \times r^{h^+h^-} (r^{K^+K^-})$.
This result has been obtained by extrapolating the DIS cross section
from $Q^2 > 0.4~{\rm GeV}^2$ ($Q^2 > 1~{\rm GeV}^2$) to $Q^2_{min}$
using a dipole form factor with a mass parameter of 10~GeV$^2$
set by the scale of the hard process.
This procedure is similar to that adopted for extracting the
open charm and $J / \Psi$ photo-production cross section
from the measured muo-production cross sections~\cite{emccc}.

Figure~\ref{fig:gluon} shows the $\eta$ distributions of the partons
(quarks and gluons) for the selected high-$p_T$ hadron and kaon pair samples.
The shapes result from the combined effect of the chosen photon energy
and the general behavior of the parton distributions.
The $\eta$ distribution is peaked around $0.1$.
For the $h^+ h^-$ sample the \qq\gq-events dominate over \qq\qqbar-events
above $\eta \sim 0.15 - 0.20$,
while for the $K^+ K^-$ sample the relative contribution of the
\qq\gq- and \qq\qqbar-events are similar above the same $\eta$.
The accessible gluon $\eta$ region extends from
$\eta \sim 0.02$ to $\eta \sim 0.2$ for a 200~GeV beam.

Assuming that in the hard scattering c.m. the directions of the selected
hadrons are approximately that of the outgoing partons we can try to
reconstruct the hard scattering kinematics using the two measured hadrons only.
Neglecting the fragmentation $p_F$, the longitudinal ($\hat{p}_L$)
and the transverse ($\hat{p}_T$) momentum components
of the {\it pseudo} jet in the partonic c.m. can be written as
\begin{equation}
\hat{p}_L = \xi \, \hat{p} \cos \hat{\vartheta}
\; \; \; \; \; \; \; \; \; \; {\rm and} \; \; \; \; \; \; \; \; \; \;
\hat{p}_T = \xi \, \hat{p} \sin \hat{\vartheta}
\label{eq:kincor}
\end{equation}
where $\hat{p} = \sqrt{\hat{s}}/2$ is the total momentum of the parton.
$\xi$ represents the energy fraction of the original parton carried away
by the hadron. 
The c.m. angle $\hat{\vartheta}$ is then related to the laboratory angle
$\vartheta_{LAB}$ between the virtual photon axis and the hadron direction by
\begin{equation}
\tan \vartheta_{LAB}^\pm = \frac{p_T}{p_L} = 
\frac{\sin \hat{\vartheta}}{\gamma \, (1 \pm \cos \hat{\vartheta})}
\end{equation}
where $\gamma = E_{\gamma^\ast} / \sqrt{\hat s}$ is the Lorentz boost factor.
The $\pm$ sign refers to $h^+$ and $h^-$, respectively.
We can also obtain the c.m. energy $\hat{s}$ 
from the measured angles $\vartheta^+_{LAB}$ and $\vartheta^-_{LAB}$
for the selected positive and negative hadron as:
\begin{equation}
\hat{s} = E_{\gamma^\ast}^2 \, \tan \vartheta^+_{LAB} \, 
                              \tan \vartheta^-_{LAB} \; .
\end{equation}
The parton momentum fraction $\eta$ follows from eqn.~\ref{eq:eta}.
The c.m. scattering angle $\hat{\vartheta}$, instead, is obtained from
\begin{equation}
\cos \hat{\vartheta} = \frac{\tan \vartheta^+_{LAB} 
                        - \, \tan \vartheta^-_{LAB}}
                     {\tan \vartheta^+_{LAB} + \, \tan \vartheta^-_{LAB}} \; .
\end{equation}

A rather good correlation between the generated momentum fraction
$\eta_{gen}$ and the reconstructed one $\eta_{rec}$
can be observed in fig.~\ref{fig:gluon}c,
which will allow us also to study $\Delta G / G$
as a function of $\eta$.
Figure~\ref{fig:gluon}d shows the correlation between the scattering c.m.
angle $\cos \hat{\vartheta}_{gen}$ and the reconstructed one
$\cos \hat{\vartheta}_{rec}$.
The generated $\cos \hat{\vartheta}_{gen}$ is defined with respect
to the outgoing quark for both the \qq\gq- and \qq\qqbar-event,
and the reconstructed $\cos \hat{\vartheta}_{rec}$
is calculated with respect to the positive hadron.
This explains the double structure of the plot,
since a quark can fragment also to a negative hadron ($\sq \rightarrow K^-$).
These correlations also demonstrate the charge retention in the process
and that the kinematics of the high-$p_T$ hadron pair is strongly correlated
with the kinematics of the outgoing partons in the c.m. (eqn.~\ref{eq:kincor}).

By selecting, for instance, hadron pairs with $|\cos \vartheta_{rec}| < 0.5$,
we will select events with $\hat{\vartheta}$ even closer to $90^\circ$
in the hard scattering c.m..
Since the Compton sub-process is peaked in the backward direction
($\cos \hat{\vartheta} \rightarrow -1$), this additional cut
will further suppress the \qq\gq-events
and therefore enhance the relative contribution of the \qq\qqbar-events.
Another possibility would be to require $x_F^- > x_F^+$,
since the Compton sub-process will preferentially generate faster positive
hadrons because of the favored fragmentation of \uq~quarks
to positive hadrons.

\section{The Results}

\begin{figure}
\vspace{-5mm}
\begin{center}
\mbox{
\epsfxsize=16cm\epsffile{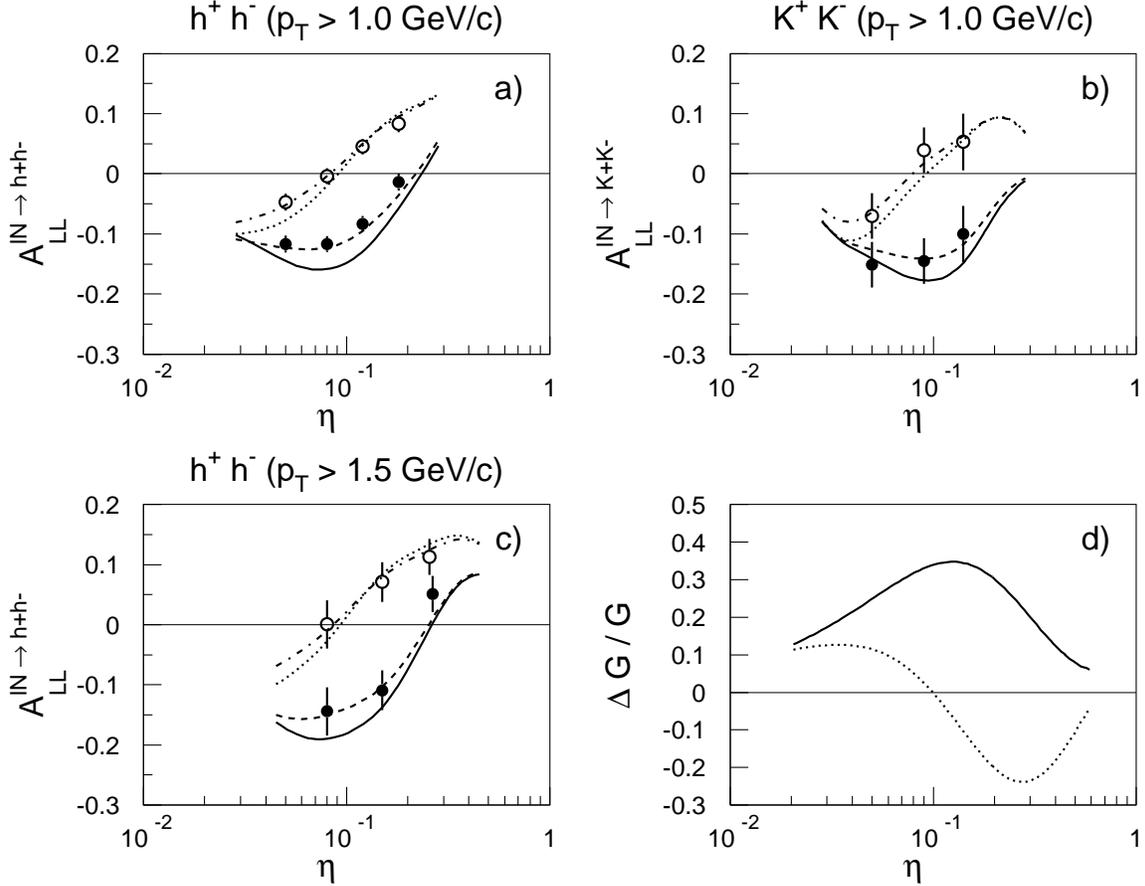}
}
\end{center}
\vspace{-10mm}
\caption{ 
Cross section spin-asymmetries $A_{LL}^{l N \rightarrow h^+ h^-}$ 
($p_T > 1~{\rm GeV}/c$) (a), $A_{LL}^{l N \rightarrow K^+ K^-}$ 
($p_T > 1~{\rm GeV}/c$) (b), and $A_{LL}^{l N \rightarrow h^+ h^-}$
($p_T > 1.5~{\rm GeV}/c$) (c) as a function of $\eta_{gen}$
with the polarized parton densities of~\protect\cite{gs96} 
(set A - full line, set C - dotted line).
The circles show the same asymmetries as a function of $\eta_{rec}$
(set A - full circles, set C - open circles)
and the error bars indicate the relative statistical precision 
for the measurement obtainable in one year.
The {\it smearing} of the asymmetry  due to the finite width
of the $\eta_{rec}$ bins (see text) is also shown
(dashed and dot-dashed lines).
Also shown are the different parametrizations for the gluon polarization
$\Delta G / G$ used at a scale of $10~{\rm GeV}^2$ (d).
}
\label{fig:all}
\end{figure}

Figure~\ref{fig:all} shows the cross section spin-asymmetries
$A_{LL}^{l N \rightarrow h^+ h^-}$ and
$A_{LL}^{l N \rightarrow K^+ K^-}$ as a function of $\eta_{gen}$
for the selected charged hadron pairs (selections {\bf A} to {\bf D}) 
and the kaon pairs (selections {\bf A} to {\bf E})
obtained by integrating eqn.~\ref{eq:all} with the POLDIS program~\cite{poldis}
and using different polarized parton densities from~\cite{gs96} (sets A and C).
The corresponding models for the gluon polarization $\Delta G / G$ 
at a scale of $10~{\rm GeV}^2$ are shown in fig.~\ref{fig:all}d.
For the set A parametrization of the polarized parton densities,
the  asymmetry $A_{LL}^{l N \rightarrow h^+ h^-}$ for the hadron pairs
(figs.~\ref{fig:all}a and~\ref{fig:all}c) is as large as $- 15~\%$
just below $\eta \sim 0.1$ and then slowly decreases, in magnitude,
with increasing $\eta$ with a zero crossing around $\eta \sim 0.2 - 0.3$.
The behavior of the asymmetry can be easily understood from the shape
of $\Delta G / G$ and the contribution of the Compton sub-process
over this $\eta$ interval:
the Compton sub-process contributes to $A_{LL}^{l N \rightarrow h^+ h^-}$
with a sign opposite to that for the PGF
and its relative contribution increases with increasing $\eta$.
For the set C the zero crossing occurs at a lower value of
$\eta$ around 0.1, reflecting the fact that $\Delta G / G < 0$ here
(fig.~\ref{fig:all}d).
The asymmetry for the selected $K^+K^-$ pairs (fig.~\ref{fig:all}b)
shows a similar behavior as a function of $\eta$.
It reaches a maximum value of almost $- 20~\%$ for $\eta$
just below 0.1 for set A.
The relative contribution of the Compton sub-process for this sample
is smaller, and the expected asymmetry is therefore larger.
The photo-production asymmetries $A_{LL}^{\gamma N \rightarrow h^+ h^-}$
and $A_{LL}^{\gamma N \rightarrow K^+ K^-}$ can be obtained from the
electro-production ones by dividing the latter ones with the depolarization
factor $D$, which is around 80~\%.

Using the correlation between the generated and reconstructed
parton momentum fraction $\eta_{gen}$ and $\eta_{rec}$ shown in
fig.~\ref{fig:gluon}c, we can divide the high-$p_T$ hadron pairs
in several $\eta_{rec}$ bins and study $A_{LL}^{l N \rightarrow h^+ h^-}$
($A_{LL}^{l N \rightarrow K^+ K^-}$) as a function of $\eta_{rec}$.
In fig.~\ref{fig:all} these asymmetries for several
$\eta_{rec}$ bins are also shown.
Because of the finite width of the correlation between $\eta_{gen}$ and
$\eta_{rec}$, each $\eta_{rec}$ bin covers a slightly larger $\eta_{gen}$
interval, which
explains the small discrepancies observed between the asymmetries
plotted as a function of $\eta_{gen}$ and $\eta_{rec}$.
This {\it smearing} of the asymmetry can be easily reproduced by
introducing a smearing in $\eta_{gen}$ for $A_{LL}^{l N \rightarrow h^+ h^-}$
($A_{LL}^{l N \rightarrow K^+ K^-}$) as shown in the same plots
(fig.~\ref{fig:all}).
Nevertheless the asymmetries calculated at the partonic level survive
the hadronization phase well, allowing one also to distinguish easily
between different parametrizations for $\Delta G / G$.

In a high rate fixed target experiment like COMPASS~\cite{comp},
integrated luminosities $L$ of 2~fb$^{-1}$ can be achieved in one year.
With the cross section
$\sigma^{l N \rightarrow h^+,h^-} = 150~{\rm nb} \times r^{h^+,h^-}$
for producing a high-$p_T$ hadron pair as discussed in the previous section,
about 700~k $h^+h^-$ pairs (cuts {\bf A} to {\bf D})
and about 70~k $K^+K^-$ pairs (cuts {\bf A} to {\bf E})
with $p_T > 1~{\rm GeV}/c$, and about 80~k $h^+h^-$ pairs with
$p_T > 1.5~{\rm GeV}/c$ can be collected.
Given a beam polarization $P_B$ of $\sim 80~\%$
and an effective target polarization $P_T$ of $\sim 25~\%$ per nucleon,
which includes the dilution factors,
these event samples will give a statistical precision in the 1~\% region 
for both $A_{LL}^{l N \rightarrow h^+h^-}$ and
$A_{LL}^{l N \rightarrow K^+K^-}$ spin-asymmetries.
Note that the measured ({\it raw}) asymmetry
$\varepsilon_{LL}^{l N \rightarrow h^+h^-} =
P_B \times P_T \times A_{LL}^{l N \rightarrow h^+h^-}$
is about $3 - 4$ times smaller.
In figure~\ref{fig:all} the expected accuracies
for the data divided in several $\eta_{rec}$ bins are also shown.

As already discussed, in order to extract $\Delta G / G$ from
$A_{LL}^{l N \rightarrow h^+h^-}$ ($A_{LL}^{l N \rightarrow K^+K^-}$)
the various backgrounds have to be subtracted from
the asymmetries (eqn.~\ref{eq:all2}).
Simulations have shown that the ratio $R = \sigma^{PGF} / \sigma^{COMPT}$
is rather stable with variations contained to within 10~\% to 15~\%
of its value for different descriptions of the fragmentation process
and various parametrizations of the unpolarized parton densities.
The polarized quark densities $\Delta {\sf q} / {\sf q}$ are already
known with a rather good precision of about 10~\%~\cite{Ade96},
and are expected to improve in the near future.
Assuming a rather generous error of 20~\% for $R$
and 10~\% for $\Delta {\sf q} / {\sf q}$,
we have estimated a statistical and systematical precision of about 5~\%
for $\Delta G / G$ from the measured
$A_{LL}^{l N \rightarrow h^+h^-}$ and $A_{LL}^{l N \rightarrow K^+K^-}$
asymmetries for each $\eta_{rec}$ bin.
The precision on the extraction of $\Delta G / G$ is mostly
affected by the Compton background subtraction,
which is larger in the $h^+h^-$ channel compared to the $K^+K^-$ one,
where, on the other hand, the event yields are smaller.
The simultaneous use of different charge and flavor combinations,
and different kinematics of the hadron pairs will give a better
determination of  $\Delta G / G$ because of the different backgrounds involved,
while the gluon densities remain the same.

\section{Conclusions}

In conclusion, we have shown that the measurement of the cross section 
spin-asymmetry for two azimuthally anti-correlated hadrons of moderate
transverse momentum $p_T$ of 1--1.5~GeV/$c$ in polarized
electroproduction on polarized targets offers a considerable sensitivity
on the gluon polarization $\Delta G / G$ at momentum fractions $\eta$
around 0.1.
Current models for $\Delta G$ predict experimental asymmetries of up
to 20~\% (5~\% with realistic dilution assumptions) and sufficiently
large cross sections to achieve statistical accuracies in the 1~\%
region for one year of running of the COMPASS experiment.

This method should be seen as complementary to the measurement of 
spin asymmetries for open charm production
discussed up to now~\cite{Glu88,comp},
since the systematic effects expected in the two processes are very different.
The PGF process not only gives opposite signs for the asymmetry of the
two processes but higher order QCD effects and hadronization effects will
be very different.
The new process offers the additional advantage that the parton kinematics
can be, at least approximately, reconstructed from the observed hadron pairs.
This will not only allow a direct measurement of the shape of 
$\Delta G (\eta)$ over the range 0.04 -- 0.2, but also checks and corrections
for model dependencies.
The systematics for the $\Delta G$ determination from the hadron pairs are
dominated by the poorly known fraction $R$ of PGF events in the total sample.
Usage of isoscalar targets reduces this to a dilution effect since the
asymmetries of the relevant backgrounds are proportional to $A_1$,
which is very small in the relevant range.

Further studies, which are beyond the scope of the present paper,
show that the simultaneous use of different selections of charge,
flavor, and kinematics will narrow down the systematic
error in the quantity $R$.
This can be seen, for instance, from figs.~\ref{fig:yields}
and~\ref{fig:gluon} which show that different selections of the hadron pairs provide us with different values for $R$
connecting the same basic quantities in eqn.~\ref{eq:all2}.

Another major uncertainty stems from the use of only leading order processes
in our estimates. We expect, however, that higher orders will not change 
the picture in a dramatic way since the relevant scales are of the order
of 10~GeV$^2$, comparable to the threshold for open charm production,
and that the polarization asymmetries survive gluon radiation through helicity
conservation.

In summary, the observation of a negative asymmetry in the proposed
measurement will be a clear signature for a non-zero
positive gluon polarization.
The experiment will be also sensitive to very small gluon polarizations,
up to five times smaller as currently discussed models.

\section*{Acknowledgments}
We would like the thank D.~De~Florian, K.~Kurek, G.K.~Mallot, and
W.~Vogelsang for valuable discussions.


\end{document}